# Systolic Arrays for Lattice-Reduction-Aided MIMO Detection

Ni-Chun Wang, Ezio Biglieri, and Kung Yao

***Abstract*—Multiple-input, multiple-output (MIMO) technology provides high data rate and enhanced QoS for wireless communications. Since the benefits from MIMO result in a heavy computational load in detectors, the design of low-complexity sub-optimum receivers is currently an active area of research. Lattice-reduction-aided detection (LRAD) has been shown to be an effective low-complexity method with near-ML performance. In this paper we advocate the use of systolic array architectures for MIMO receivers, and in particular we exhibit one of them based on LRAD. The "LLL lattice reduction algorithm" and the ensuing linear detections or successive spatial-interference cancellations can be located in the same array, which is considerably hardware-efficient. Since the conventional form of the LLL algorithm is not immediately suitable for parallel processing, two modified LLL algorithms are considered here for the systolic array. LLL algorithm with full-size reduction (FSR-LLL) is one of the versions more suitable for parallel processing. Another variant is the all-swap lattice-reduction (ASLR) algorithm for complex-valued lattices, which processes all lattice basis vectors simultaneously within one iteration. Our novel systolic array can operate both algorithms with different external logic controls. In order to simplify the systolic array design, we replace the Lovász condition in the definition of LLL-reduced lattice with the looser Siegel condition. Simulation results show that for LR-aided linear detections, the bit-error-rate performance is still maintained with this relaxation. Comparisons between the two algorithms in terms of bit-error-rate performance, and average FPGA processing time in the systolic array are made, which shows that ASLR is a better choice for a systolic architecture, especially for systems with a large number of antennas.**

## I. Introduction

MULTIPLE-INPUT, multiple-output (MIMO) technology, using several transmit and receive antennas in a rich-scattering wireless channel, has been shown to provide considerable improvement in spectral efficiency and channel capacity [1]. MIMO systems yield spatial diversity gain, spatial multiplexing gain, array gain, and interference reduction over single-input single-output (SISO) systems [2]. However, these benefits come at the price of a computational complexity of the detector that may be intolerably large. In fact, optimal maximum-likelihood (ML) detection in large MIMO systems may not be feasible in real-time applications as its complexity increases exponentially with the number of antennas. Low-complexity receivers, employing linear detection or successive spatial-interference cancellation (SIC), are computationally less heavy, and amenable to simple hardware implementation [3]–[5]. However, diversity and error-rate performance of these low-complexity detectors are not comparable to those achieved with ML.

Lattice-reduction-aided detection (LRAD), which combines lattice reduction techniques with linear detections or SIC, has been shown to yield some improvement on error-rate performance [6]–[8]. Lenstra-Lenstra-Lovász (LLL) algorithm [9] is the most widely used lattice reduction algorithm, and can be applied to complex-valued lattices [10]. The performance of complex LLL-aided linear detection in MIMO systems was analyzed in [11]. LLL-based LRAD was also shown to achieve full receiver diversity [12]. It was also shown that the LR-aided minimum mean-square-error decoding achieves the optimal diversity-multiplexing tradeoff [16]. When applied to MIMO detection, the average complexity of LLL algorithm is polynomial in the dimension of the channel matrix (the worst-case complexity could be unbounded [13]). A fixed-complexity LLL algorithm, which modifies the original version to allow more robust early termination, has recently been proposed in [17]. In LRAD, LLL algorithm need be performed only when the channel state changes. If the channel change rate is high, or a large number of channel matrices need be processed such as in a MIMO-OFDM system, a fast-throughput algorithm and the corresponding implementation structure is needed for real-time applications. To obtain this, we first discuss two variants of LLL algorithm, suitably modified for parallel processing. Second, we propose a novel systolic array structure implementing the two modified LLL algorithms and the ensuing detection methods.

A systolic array [18], [19] is a network of processing elements (PE) which transfer data locally and regularly with nearby elements and work rhythmically. In Fig. 1(a), a simple two-dimensional systolic array is shown as an example. In this case, the matrix operation $\mathbf{D} = \mathbf{A} \cdot \mathbf{B} + \mathbf{C}$ is calculated by the systolic array, where $\mathbf{A}$, $\mathbf{B}$, $\mathbf{C}$ and $\mathbf{D}$ are $2 \times 2$ matrices. The operation of each PE is shown in Fig. 1(b). The inputs of the systolic array, the entries of matrices $\mathbf{A}$ and $\mathbf{C}$, are pipelined in a slanted manner for proper timing. Since all PEs can work simultaneously, the latency is shorter than with a single processor system, and the results of $\mathbf{D}$ are outputted in parallel. Systolic algorithms and the corre-

N.C. Wang, E. Biglieri, and K. Yao are with the Electrical Engineering Department, University of California-Los Angeles, Los Angeles, CA 90095, USA (Address: 56-125B Engineering IV Building, 420 Westwood Plaza, Los Angeles, CA 90095, USA; e-mail: nichun@ee.ucla.edu; e.biglieri@ieee.org; yao@ee.ucla.edu). The work of N.C. Wang was partially supported by National Science Council, Taiwan (R.O.C.). (TMS-094-2-A-002). The work of K. Yao was partially supported by NSF CENS program CCR-012, NSF grant EF-0410438, and NSF grant DBI-0754247.

E. Biglieri is also with the Departament de Tecnologies de la Informació i les Comunicacions, Universitat Pompeu Fabra, Barcelona, Spain. The work of E.B. was supported by the Spanish Ministry of Education and Science under Project CONSOLIDER-INGENIO 2010 CSD2008-00010 "COMONSENS".

sponding systolic arrays have been designed for a number of linear algebra algorithms, such as matrix triangularization [20], matrix inversion [21] , adaptive nulling [22], recursive least-square [23], [24], etc. An overview of systolic designs for several computationally demanding linear algebra algorithms for signal processing and communications applications was recently published in [25]. While systolic arrays allow simple parallel processing and achieve higher data rates without the demand on faster hardware capabilities, the existence of multiple PEs implies a higher cost of circuit area. Thus, time efficiency is traded off with circuit area in hardware design. For the application we are advocating in this paper (MIMO detectors), systolic arrays offer an attractive solution, as we must cope with a high computational load while requiring high throughput and real-time operation. Systolic arrays have been previously suggested for MIMO applications. In [26], the authors proposed a universal systolic array for adaptive and conventional linear MIMO detectors. In [27], a reconfigurable systolic array processor based on coordinate rotation digital computer (CORDIC) [28] is proposed to provide efficient MIMO-OFDM baseband processing. Also, matrix factorization and inversion are widely used in MIMO detection, with systolic arrays used to increase the throughput [5], [29].

In this paper, our objective is to provide a novel systolic array design for LLL-based LRAD. The ideas are described from a system-level perspective instead of detailed discussion on the hardware-oriented issues. The system model and how LRAD works are briefly described in Section II. Since the original LLL algorithm [8]–[15] is not designed for parallel processing, and hence is not suitable for systolic design, two modified LLL algorithms are considered here (Section III). Note that we are not claiming the two algorithms works better than the original LLL in terms of the LRAD bit-error-rate (BER) performance. First, we improve on the format of conventional LLL algorithm by altering the flow of size-reduction process (we call it "LLL with full size-reduction," or FSR-LLL). FSR-LLL is more time-efficient in parallel processing than the conventional format, and hence suitable for systolic design. We also consider a variant of the LLL algorithm called "all-swap lattice reduction (ASLR)," which was first proposed in [30] for real lattices, and derive its complex-number version. A crucial difference between ASLR and LLL algorithm is that with ASLR all lattice basis vectors are simultaneously processed during a single iteration. In both algorithms, in order to simplify the systolic array operations we replace the original Lovász condition [9] of LLL algorithm with the slightly weaker Siegel condition [31]. Surprisingly, for LR-aided linear detections the BER performance with Siegel condition under the proper parameter setting is just as good as the one using Lovász condition. However, for LR-aided SIC, the performance with Lovász condition is still slightly better due to less error propagation. The mapping from algorithm to systolic array is introduced in Section IV. In our design, ASLR and FSR-LLL can be operated on the same systolic-array structure, but the external logic controller is also required to control the algorithm flow. Additionally, since ASLR was originally designed for parallel processing, a systolic array running ASLR is on the average more efficient than one running FSR-LLL. Simulation results also show that ASLR-based LRAD has a BER performance very similar to that of LLL algorithm. Comparison between our proposed design and the conventional LLL in FPGA implementation shows that the systolic arrays do provide faster processing speed with a moderate increase of hardware resources. After the channel state matrix has been lattice-reduced, linear detectors or SIC can also be implemented by the same systolic array without any extra hardware cost, which is discussed in Section V.

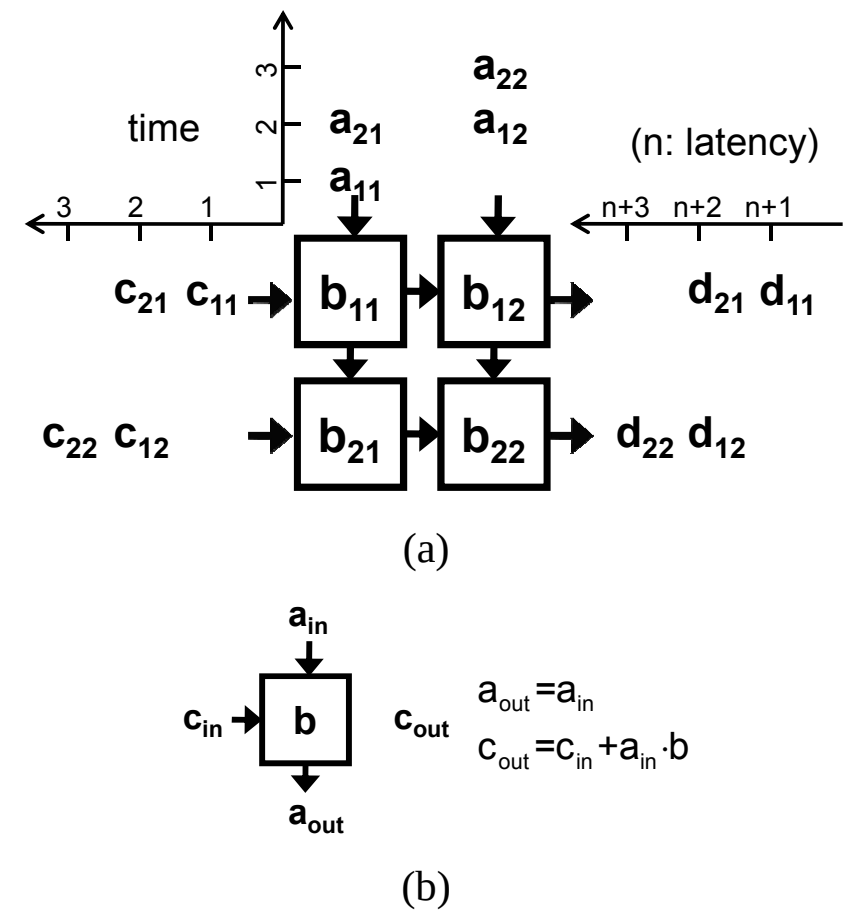

Fig. 1. (a) Two-dimensional systolic array performing matrix calculation $\mathbf{D} = \mathbf{A} \cdot \mathbf{B} + \mathbf{C}$ , where $a_{ij}, b_{ij}, c_{ij}, d_{ij}$ are the $(i, j)$ entries of the matrix $\mathbf{A}, \mathbf{B}, \mathbf{C}$, and $\mathbf{D}$. (b) The operation of each processing element.

The following notations are used throughout the remaining sections. Capital bold letters denote matrices, and lower case bold letters denote column vectors. For example, $\mathbf{X} = [\mathbf{x}_1, \mathbf{x}_{2,} \cdots, \mathbf{x}_m]$ is a matrix with $m$ columns of $\mathbf{x}_1$ to $\mathbf{x}_m$. The entry of a matrix $\mathbf{X}$ at position $(i, j)$ is denoted by $x_{i,j}$, and the $k^{th}$ element of a vector $\mathbf{x}$ is denoted by $x_k$. The submatrix (subvector) formed from the $a^{th}$ to $b^{th}$ rows and $m^{th}$ to $n^{th}$ columns of $\mathbf{X}$ is denoted by $\mathbf{X}_{a:b,m:n}$. The notations $(\cdot)^+$, $(\cdot)^T$, $(\cdot)^H$ and $(\cdot)^\dagger$ are used for conjugate, transpose, Hermitian transpose, and Moore-Penrose pseudo-inverse of a matrix, respectively. $\|\mathbf{x}\|$ is the Euclidean norm of the vector $\mathbf{x}$. $\Re(\cdot)$ and $\Im(\cdot)$ are the real and imaginary parts of a complex number, respectively. $\lceil x \rfloor$ indicates the closest integer to $x$. If $x$ is a complex number, then $\lceil x \rfloor = \lceil \Re(x) \rfloor + i \lceil \Im(x) \rfloor$. $\mathbf{I}_m$ and $\mathbf{0}_m$ are $m \times m$ identity and null matrices, respectively.

## II. Lattice-Reduction-Aided Detection

### A. System Model

We consider a MIMO system with $m$ transmit and $n$ receive antennas in a rich-scattering flat-fading channel. Spatial multiplexing is employed, so that data are transmitted as $m$ substreams of equal rate. These substreams are mapped onto M-ary QAM symbols. Let $\mathbf{x}$ denote the complex-valued $m \times 1$ transmitted signal vector, and $\mathbf{y}$ the complex-valued $n \times 1$ received signal vector. The baseband model for this MIMO system is

$$\mathbf{y} = \mathbf{H}\mathbf{x} + \mathbf{n}, \tag{1}$$

where $\mathbf{H}$ is the $n \times m$ channel matrix: its entries are uncorrelated, zero-mean, unit-variance complex circularly symmetric

Gaussian fading gains $h_{ij}$, and $\mathbf{n}$ is the $n \times 1$ additive white complex Gaussian noise vector with zero mean and $E[\mathbf{n}\mathbf{n}^H] = \sigma^2\mathbf{I}$. The average power of each transmitted signal $x_i$ is assumed to be normalized to 1, i.e., $E[\mathbf{x}\mathbf{x}^H] = \mathbf{I}$. Additionally, we assume that the channel matrix entries are fixed during each frame interval, and the receiver has perfect knowledge of the realization of $\mathbf{H}$.

### *B. Linear Detection*

In linear detection, the estimated signal $\hat{\mathbf{x}}$ is computed by first premultiplying the received signal $\mathbf{y}$ by an $n \times m$ "weight matrix" $\mathbf{W}$. The two most common design criteria for $\mathbf{W}$ are zero-forcing (ZF) and minimum mean-square error (MMSE). In zero-forcing detection, the weight matrix $\mathbf{W}_{ZF}$ is set to be the Moore-Penrose pseudo-inverse $\mathbf{H}^\dagger$ of the channel matrix $\mathbf{H}$, i.e.,

$$\hat{\mathbf{x}}_{ZF} = \mathbf{W}_{ZF}\mathbf{y} = \mathbf{H}^\dagger\mathbf{y} = \mathbf{x} + \mathbf{H}^\dagger\mathbf{n}. \quad (2)$$

It is known that zero-forcing detection suffers from the noise enhancement problem, as the channel matrix may be ill-conditioned. Under the MMSE criterion, the weight matrix $\mathbf{W}$ is chosen in such a way that the mean-squared-error between the transmitted signal $\mathbf{x}$ and the estimated signal $\hat{\mathbf{x}}$ is minimized. The mean-squared-error (MSE) is defined as $MSE \triangleq E[\|\mathbf{x} - \hat{\mathbf{x}}\|^2] = E\left[(\mathbf{x} - \mathbf{W}\mathbf{y})^H(\mathbf{x} - \mathbf{W}\mathbf{y})\right]$. The weight matrix $\mathbf{W}$ that minimizes the MSE is

$$\mathbf{W}_{MMSE} = (\mathbf{H}^H\mathbf{H} + \sigma^2\mathbf{I})^{-1}\mathbf{H}^H, \quad (3)$$

It is well known that, as $\sigma^2 \to 0$, the weight matrix $\mathbf{W}_{MMSE}$ approaches $\mathbf{W}_{ZF}$. Since $\mathbf{W}_{MMSE}$ takes noise power into consideration, MMSE detection suffers less from noise enhancement than ZF detection. In [8], [32], it is shown that MMSE is equivalent to ZF in an extended system model, i.e.,

$$\hat{\mathbf{x}}_{MMSE} = \mathbf{W}_{MMSE}\underline{\mathbf{y}} = \underline{\mathbf{H}}^\dagger\underline{\mathbf{y}} = (\underline{\mathbf{H}}^H\underline{\mathbf{H}})^{-1}\underline{\mathbf{H}}^H\underline{\mathbf{y}}, \quad (4)$$

where

$$\underline{\mathbf{H}} = \begin{bmatrix} \mathbf{H} \\ \sigma\mathbf{I}_m \end{bmatrix} \text{and} \underline{\mathbf{y}} = \begin{bmatrix} \mathbf{y} \\ \mathbf{0}_{m\times 1} \end{bmatrix}. \quad (5)$$

Comparing (2) with (4), it follows that the two detection methods can share the same structure in systolic-array implementation, which we shall elaborate upon in Section IV.

### *C. Lattice-Reduction-Aided Linear Detection*

The idea underlying lattice reduction is the selection of a basis vector for the lattice under some goodness criterion [33]. We first observe that, under the assumption of QAM transmission, the transmitted vector $\mathbf{x}$ is an integer point of a square lattice (after proper scaling and shifting of the original QAM constellation). By interpreting the columns of the channel matrix $\mathbf{H}$ as a set of lattice basis vectors, $\mathbf{H}\mathbf{x}$ is also a lattice point. If two basis sets $\mathbf{H}$ and $\tilde{\mathbf{H}}$ are related by $\tilde{\mathbf{H}} = \mathbf{H} \cdot \mathbf{T}$, $\mathbf{T}$ a *unimodular* matrix, they generate the same set of lattice points. In MIMO detection, the objective of the lattice reduction algorithm is to derive a better-conditioned channel matrix $\tilde{\mathbf{H}}$. In this paper, we focus on the complex-valued LLL algorithm [10], [11]. More details about the LLL algorithm will be provided in Section III.

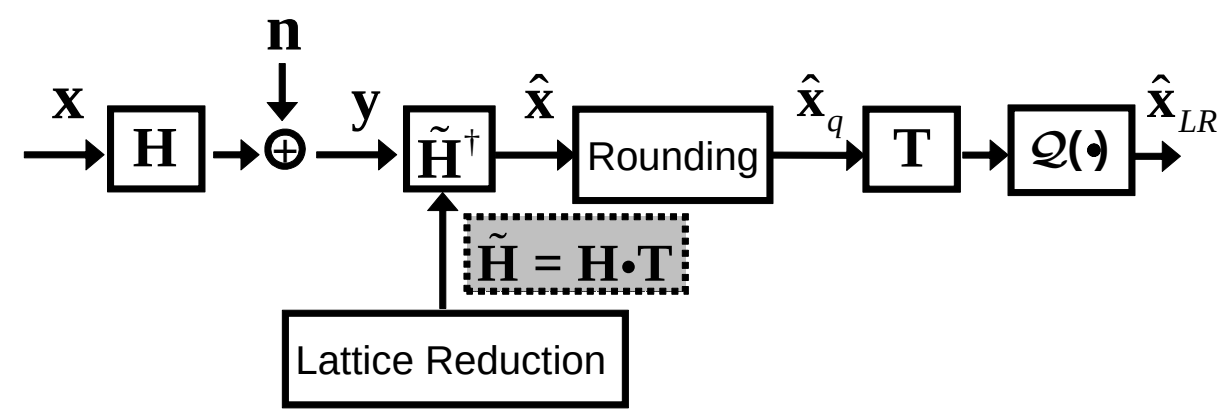

Fig. 2. Block diagram of linear lattice-reduction-aided detection

After lattice-reduction of the channel matrix, we can perform the linear detection, as described in Section II-B, based on $\tilde{\mathbf{H}}$. Consider ZF first. The estimated signal $\hat{\mathbf{x}}$ can be written as

$$\hat{\mathbf{x}} = \tilde{\mathbf{H}}^\dagger\mathbf{y} = \tilde{\mathbf{H}}^\dagger\left((\mathbf{H}\mathbf{T})(\mathbf{T}^{-1}\mathbf{x}) + \mathbf{n}\right) = \mathbf{T}^{-1}\mathbf{x} + \tilde{\mathbf{H}}^\dagger\mathbf{n}. \quad (6)$$

Since $\hat{\mathbf{x}}$ is no longer an integer vector, the simplest but suboptimal way of estimating $\mathbf{T}^{-1}\mathbf{x}$ is to round $\hat{\mathbf{x}}$ element-wise to the nearest integer. Let $\hat{\mathbf{x}}_q$ be an estimate of $\mathbf{T}^{-1}\mathbf{x}$ after rounding. The final step is to transform $\hat{\mathbf{x}}_q$ back into an estimate of $\mathbf{x}$, which is done by multiplying $\hat{\mathbf{x}}_q$ by the unimodular matrix $\mathbf{T}$. Since the vector entries after the transformation could lie outside the QAM constellation boundary, we finally quantize those points outside the boundary to the closest constellation point, i.e., $\hat{\mathbf{x}}_{LR} = \mathcal{Q}(\mathbf{T}\hat{\mathbf{x}}_q)$. Fig. 2 shows the block diagram of LR-aided ZF detection for MIMO. It is easy to see that the same structure can also be used for MMSE detection, by simply replacing $\mathbf{H}$ and $\mathbf{y}$ with the extended matrix $\underline{\mathbf{H}}$ and the vector $\underline{\mathbf{y}}$ defined in (5), respectively. The remaining operations are the same as in ZF.

### *D. LR-Aided Successive Spatial-Interference Cancellation*

Besides being suitable linear detection systolic design can be used to exploit the regularity of successive spatial-interference cancellation (SIC). In [8], it is shown that LR-aided SIC outperforms linear detection methods, while exhibiting a complexity comparable to linear detection. The LR-aided SIC can be conveniently described in terms of the QR decomposition of the reduced channel matrix. Here we summarize briefly the procedure of LR-aided ZF-SIC only, as the LR-aided MMSE-SIC can be derived in a similar way. Let the QR decomposition of the reduced channel matrix be $\tilde{\mathbf{H}} = \tilde{\mathbf{Q}}\tilde{\mathbf{R}}$. First, multiply $\tilde{\mathbf{Q}}^H$ to $\mathbf{y}$ in (1), we obtain

$$\mathbf{v} \triangleq \tilde{\mathbf{Q}}^H\mathbf{y} = \tilde{\mathbf{R}}\mathbf{z} + \tilde{\mathbf{Q}}^H\mathbf{n}, \text{ where } \mathbf{z} = \mathbf{T}^{-1}\mathbf{x}. \quad (7)$$

Then we can solve for $\mathbf{z}$ layer by layer starting from the bottom to the top, i.e.

$$\hat{z}_i = \left\lceil \frac{\mathbf{v}_i}{\tilde{r}_{ii}} \right\rfloor, \quad \mathbf{v} := \mathbf{v} - (\tilde{\mathbf{R}}_{1:i,i})\hat{z}_i, \quad (8)$$

where $i$ starts from $m$ to 1 and $\hat{z}_i$ is the estimate of each entry of $\mathbf{z}$.

## III. TWO VARIANTS OF LLL ALGORITHM

In this section, we introduce two variants of LLL algorithm which are more time-efficient than the classical LLL algorithm when using parallel processing. Since systolic arrays yield a simple form of parallel processing, our systolic array design for LRAD is based on these two algorithms.

We begin the discussion with the definition of LLL-reduced lattice. Let $\mathbf{H}$ (an $n \times m$ matrix) be a set of lattice basis vectors, with QR decomposition $\mathbf{H} = \mathbf{QR}$. The basis set $\mathbf{H}$ is *complex LLL-reduced* with parameter $\delta$ $(1/2 < \delta < 1)$, if the following two conditions are satisfied [10], [11]:

(a)

$$\mu_{i,j} \triangleq \frac{r_{i,j}}{r_{i,i}}, \qquad |\Re(\mu_{i,j})| \leq \tfrac{1}{2} \text{ and } |\Im(\mu_{i,j})| \leq \tfrac{1}{2}, \quad 1 \leq i < j \leq m, \tag{9}$$

(b)

$$\delta - \left| \frac{r_{i-1,i}}{r_{i-1,i-1}} \right|^2 \leq \frac{|r_{i,i}|^2}{|r_{i-1,i-1}|^2}, \; 2 \leq i \leq m. \tag{10}$$

The second condition in (10) is called the *Lovász condition*, and the process to make the basis set satisfy (9) is called *size reduction*. In the standard form of LLL algorithm considered in the literature [8]–[15], size reduction applies only to one column of $\mathbf{H}$ during a single iteration. Now, systolic arrays, allowing simple parallel processing, are capable of updating the whole matrix without introducing extra delays. Hence, our proposed systolic array is first designed based on the LLL algorithm in a different form, which we call it "LLL algorithm with full size reduction (FSR-LLL)."

### A. LLL algorithm with Full Size Reduction (FSR-LLL)

Table I shows the LLL algorithm with full size reduction. In the following discussion, we refer to the lines in Table I. There are three main differences between FSR-LLL and the conventional complex LLL algorithm[1], although the lattice reduced bases from both algorithms are still the same. First, the full size reduction (lines 4~10) is executed in each iteration of the *while* loop (line 3), which means that all columns of $\mathbf{R}$ and $\mathbf{T}$ are size-reduced at the beginning of each iteration. The advantage here is that, once condition (10) is also fulfilled after full size reduction (i.e., no $k'$ is found in line 11), then the FSR-LLL can immediately end the process (line 20). For example, suppose that $k$ equals 3 at current iteration. Since all columns in $\mathbf{R}$ and $\mathbf{T}$ are size-reduced after full size reduction, if no $k'$ can be found in line 11 (a search that a systolic array can make in parallel), then no further processing is needed in FSR-LLL. However, in the conventional LLL format, the process will end until columns 3 to $m$ are sequentially size-reduced. With a systolic-array implementation, FSR-LLL is faster, and its efficiency is especially apparent when $m$ is large. The second difference is that the Givens rotation (lines 13~16) is executed before the column swap (line 17). This is because the Givens rotation process can work in parallel with full size reduction, whereas the columns swap cannot. This point will be made clear in Section IV-A. Third, the QR decomposition $\mathbf{Q}^H\mathbf{H} = \mathbf{R}$ is considered as the input of the algorithm, instead of $\mathbf{H} = \mathbf{QR}$. From line 16, the Givens rotation matrix $\mathbf{G}$ applies to the same two rows of $\mathbf{Q}^H$ and $\mathbf{R}$, which simplifies the design of the systolic array. Additionally, after FSR-LLL, $\tilde{\mathbf{Q}}^H$ is ready for calculating the pseudoinverse of $\tilde{\mathbf{H}}$ for linear-detection.

[1]For comparison, the interested readers can refer to the Table I in [11] for the conventional complex LLL algorithm. The Table I and II in this paper are presented in the similar format as the one in [11]. All the simulation results related to the conventional LLL in this paper are also based on the same table.

TABLE I
LLL ALGORITHM WITH FULL SIZE REDUCTION

*INPUT* $\mathbf{Q}^H, \mathbf{R}$
*OUTPUT* $\tilde{\mathbf{Q}}^H = \mathbf{Q}^H, \tilde{\mathbf{R}} = \mathbf{R}, \mathbf{T}$
(1) Initialization $\mathbf{T} = \mathbf{I}_m$
(2) $k = 2$
(3) While $k \leq m$
**Full Size Reduction**
(4) for $j = m, \cdots, 2$
(5) for $i = j-1, \cdots, 1$
(6) $\mu_{i,j} = [\![ r_{i,j}/r_{i,i} ]\!]$
(7) $\mathbf{R}_{1:i,j} := \mathbf{R}_{1:i,j} - \mu_{i,j}\mathbf{R}_{1:i,i}$
(8) $\mathbf{T}_{1:m,j} := \mathbf{T}_{1:m,j} - \mu_{i,j}\mathbf{T}_{1:m,i}$
(9) end
(10) end
(11) Find the smallest $k'$ between $k \sim m$ such that $\delta - |r_{k'-1,k'}/r_{k'-1,k'-1}|^2 > |r_{k',k'}|^2 / |r_{k'-1,k'-1}|^2$ :
(12) If $k'$ exists
**Givens Rotation**
(13) $\eta_1 = r_{k'-1,k'} / \|r_{k'-1:k',k'}\|$
(14) $\eta_2 = r_{k',k'} / \|r_{k'-1:k',k'}\|$
(15) $\mathbf{G} = \begin{bmatrix} \eta_1^+ & \eta_2 \\ -\eta_2 & \eta_1 \end{bmatrix}$
(16) $\mathbf{R}_{k'-1:k',k'-1:m} := \mathbf{G} \cdot \mathbf{R}_{k'-1:k',k'-1:m}$, $\mathbf{Q}^H_{k'-1:k',1:n} := \mathbf{G} \cdot \mathbf{Q}^H_{k'-1:k',1:n}$
**Column Swap**
(17) Swap columns $k'-1$ and $k'$ in $\mathbf{R}$ and $\mathbf{T}$
(18) $k := \max\{k'-1, 2\}$
(19) else
(20) $k := m+1$
(21) end
(22) end

### B. All-Swap Lattice Reduction (ASLR) Algorithm

The ASLR algorithm is a variant of the LLL algorithm, and was first proposed for real number lattices only [30]. Table II describes its extension to a complex version. One significant difference between FSR-LLL and ASLR is that every pair of columns $k$ and $k-1$ with even (or odd) index $k$ could be swapped simultaneously. The algorithm begins with full size reduction, which is the same as FSR-LLL. Givens-rotation and column-swap operations (same as in Table I, lines 13~17) should be executed on all possible even (odd) $k$ that violate the condition in (10), and then start another iteration with the indicator variable "*order*" set to odd (even). If condition (10) holds for all even (odd) $k$, Givens rotation and columns swap will not be executed. Meanwhile, we can immediately check for all odd (even) $k$ instead. Matrix $\mathbf{R}$ is already full-size reduced, with no need to start the next iteration with full size

TABLE II
ALL SWAP LATTICE REDUCTION ALGORITHM

*INPUT* $\mathbf{Q}^H, \mathbf{R}$
*OUTPUT* $\tilde{\mathbf{Q}}^H = \mathbf{Q}^H, \tilde{\mathbf{R}} = \mathbf{R}, \mathbf{T}$
(1) Initialization $\mathbf{T} = \mathbf{I}_m$
(2) *order*=EVEN
(3) While (any swap is possible in lines (9) or (16) )
**Full Size Reduction**
(4) Execute lines 4 ~ 10 in Table I
**Givens Rotation and Column Swap**
(5) If *order*=EVEN
(6) If $\delta - \left|r_{k-1,k}/r_{k-1,k-1}\right|^2 \leq \left|r_{k,k}\right|^2 / \left|r_{k-1,k-1}\right|^2$ for all even $k$
(7) go to line (13)
(8) else
(9) Execute lines 13~17 in Table I for all even $k$ between $2 \sim m$ such that $\delta - \left|\mu_{k-1,k}\right|^2 > \left|r_{k,k}\right|^2 / \left|r_{k-1,k-1}\right|^2$
(10) *order* = ODD
(11) end
(12) else
(13) If $\delta - \left|r_{k-1,k}/r_{k-1,k-1}\right|^2 \leq \left|r_{k,k}\right|^2 / \left|r_{k-1,k-1}\right|^2$ for all odd $k$
(14) go to line (6)
(15) else
(16) Execute lines 13~17 in Table I for all odd $k$ between $2 \sim m$ such that $\delta - \left|\mu_{k-1,k}\right|^2 > \left|r_{k,k}\right|^2 / \left|r_{k-1,k-1}\right|^2$
(17) *order*=EVEN
(18) end
(19) end
(20) end

reduction (Table II, line 7 or 14). If neither an even nor odd $k$ violates condition (10) after full size reduction, the ASLR process ends.

### C. Replacing Lovász condition with Siegel condition

From the previous discussion, it is clear that all basis vectors are size reduced within one processing iteration of full size reduction. Additionally, according to line 11 in Table I and lines 6 and 13 in Table II, the lattices processed by FSR-LLL and ASLR both satisfy the Lovász condition in (10). Therefore, we can conclude that these two algorithms also generate LLL-reduced lattice. Consequently, like the conventional LLL, FSR-LLL-aided and ASLR-aided detection also achieves full receive diversity in MIMO system [11], [12].

The Lovász condition involves two diagonal elements and one off-diagonal element in the matrix $\mathbf{R}$. In order to simplify the data communication between processing elements in the systolic array, we relax the Lovász condition by replacing it with

$$\delta - \frac{1}{2} \leq \frac{\left|r_{i,i}\right|^2}{\left|r_{i-1,i-1}\right|^2},\ 2 \leq i \leq m, \tag{11}$$

where $\delta$ lies in the range $(1/2, 1)$ , the same as for Lovász condition. The condition (11) is also called Siegel condition [31], and it is weaker than the Lovász condition because

$$\delta - \frac{1}{2} \leq \delta - \left|\frac{r_{i-1,i}}{r_{i-1,i-1}}\right|^2 \leq \frac{\left|r_{i,i}\right|^2}{\left|r_{i-1,i-1}\right|^2},\ 2 \leq i \leq m. \tag{12}$$

The first inequality follows from (9). Similar approximation as in (11) can be found in [34]. The advantage of using this new condition is that only two neighboring diagonal elements of $\mathbf{R}$ are involved. We will have more discussion on the impact of designing systolic array with this new condition in Section IV. Another advantage comes from the fact that the new condition check can be done by taking the square-root in (11). In hardware implementation, it implies that we can save precision bits by storing $|r_{i,i}|/|r_{i-1,i-1}|$ rather than $|r_{i,i}|^2 \big/ |r_{i-1,i-1}|^2$. Additionally, the condition check can be done without a division, simply by comparing the value of $|r_{i,i}|$ and $\sqrt{\delta - 1/2}\,|r_{i-1,i-1}|$ , where $\sqrt{\delta - 1/2}$ is a pre-computed constant once $\delta$ is determined. In the balance of this paper, when we refer to FSR-LLL and ASLR we mean FSR-LLL and ASLR *with Siegel condition*.

Since Siegel condition is weaker than Lovász condition, one might expect the performance of the lattice reduction algorithm with condition (11) to be worsened. Yet, by a proof similar to that in [11], [12] we can show that the LLL algorithm with Siegel condition also achieves maximum receive diversity in MIMO systems. In the proof of LLL-aided detection achieving full diversity [11], [12], the key step and the only step involving the LLL-reduced conditions is that the orthogonality defect $\kappa$ ($\kappa \geq 1$) of the LLL-reduced basis set $\mathbf{H}$ is upper bounded by

$$\kappa \triangleq \frac{\prod_{i=1}^{m} \|\mathbf{h}_i\|^2}{\det\left(\mathbf{H}^H\mathbf{H}\right)} \leq 2^{-m}\left(\frac{2}{2\delta - 1}\right)^{\frac{m(m+1)}{2}}, \tag{13}$$

where $\mathbf{h}_i$'s are the columns of $\mathbf{H}$. In particular, (13) also holds for the lattices reduced by LLL algorithm with Siegel condition. This can be justified by the same proof as in [11, Appendix B], whose details will be omitted in this paper. Hence, the LLL algorithm with the Lovász condition replaced by the Siegel condition also achieves maximum diversity in MIMO system. However, achieving maximum receive diversity does not automatically imply that the bit-error-rate (BER) performance is as good as using the conventional LLL algorithm. One can easily observe that if $\delta$ is very close to $1/2$ , condition (11) is almost always true. Thus, the Givens rotation and column swap steps in the reduction algorithm would seldom be performed, which causes the BER performance to be much worse than with conventional LLL. On the contrary, as $\delta$ approaches $1$ one can expect the performance of FSR-LLL and ASLR to be closer to the conventional LLL. In Fig. 3, we show the empirical cumulative probability functions of the orthogonality defect $\kappa$ for $4 \times 4$ channel matrices under three different reduction algorithms. The results of FSR-LLL and ASLR overlap for all three values of $\delta$, which implies that the effects of these two method on lattice reduction are almost the same. As $\delta = 0.99$, FSR-LLL and ASLR give a result close to the LLL with $\delta = 0.75$, which is a very common setting as documented in previous works [8], [9], [12]. For $\delta = 0.51$ and $0.75$, the gap between LLL and FSR-LLL (ASLR) is much larger than for $\delta = 0.99$. In section IV-C, we will show that for $\delta$ equal to $0.99$, the BER performance of LR-aided linear detections using FSR-LLL and ASLR is not worse than the one using the conventional LLL with the same $\delta$ value. Based on these results, in our systolic array design we choose $\delta = 0.99$.

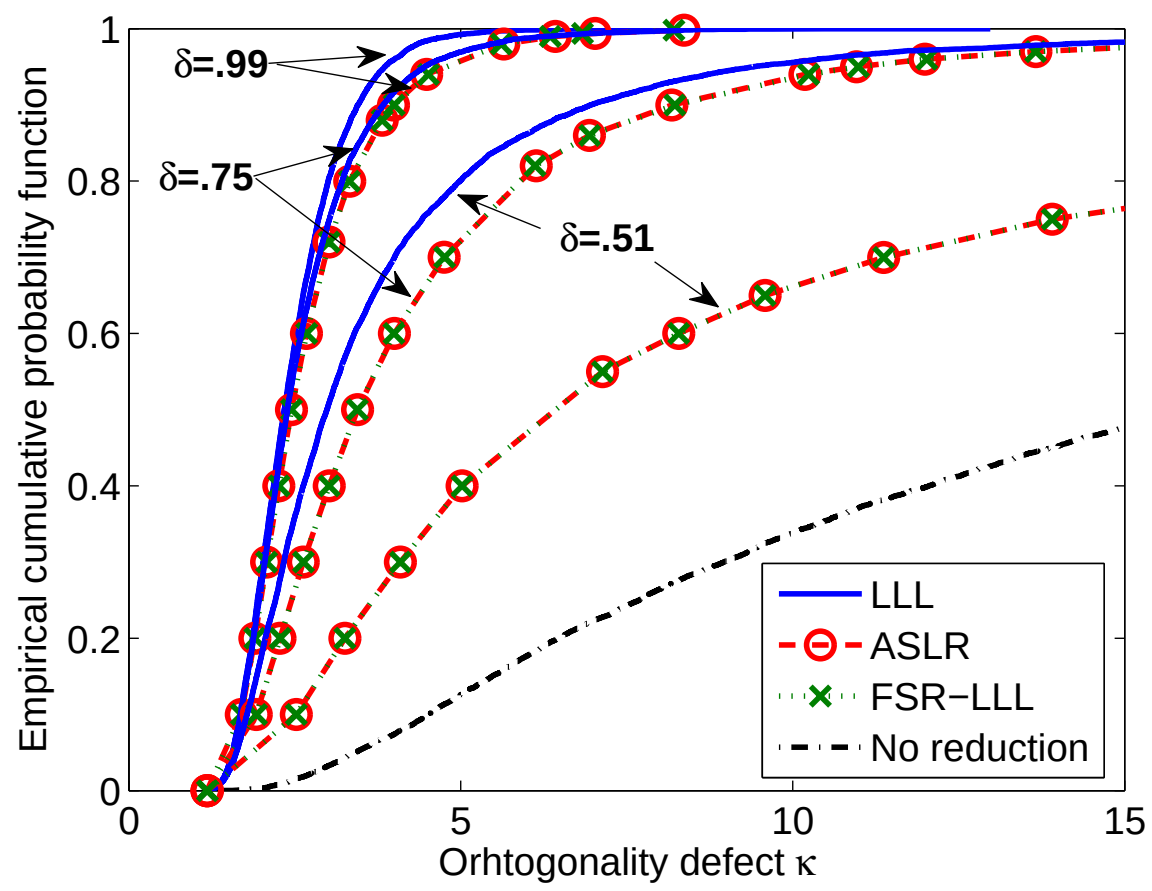

Fig. 3. The empirical cumulative probability functions of the orthogonality defect $\kappa$ for the $4 \times 4$ channel matrices under three different reduction algorithms.

## IV. Systolic Array for Two Lattice-Reduction Algorithms

From Fig. 2, the whole process of LRAD can be viewed as taking two steps: lattice reduction for the channel matrix, and detection. In this section, we exhibit our systolic array design for LLL lattice reduction algorithm. The ensuing linear detection or SIC on systolic array will be discussed in Section V. In the following discussion, we assume that the channel matrix has been QR decomposed. It is known that QRD can be implemented in systolic array based on a series of Givens rotations, since Given rotations can be executed in a parallel manner [20]–[22]. Since the conventional systolic array for QRD usually contains square root operations, which are computationally intensive in hardware implementation, a square-root-free systolic QRD based on Squared Givens rotations (SGR) can be used (the interested readers can refer to [29], [35]). In [8], it is also shown that the sorted QRD (SQRD) can reduce the number of column swaps in the LLL algorithm, and hence leads to less processing time. However, it also requires higher hardware complexity and latency to implement SQRD than the conventional QRD [36].

### A. Systolic Array for FSR-LLL

In the following, we assume a $4 \times 4$ MIMO system (i.e., $m = 4$, $n = 4$) and illustrate the proposed systolic algorithm in three parts: full size reduction, Givens rotation, and column swap.

*1) Full Size Reduction:* The systolic array for the remaining parts of LRAD is shown in Fig. 4(a) . Four different kinds of PEs are used, viz., diagonal cells, off-diagonal cells, vectoring cells, and rotation cells. For the full size reduction part, only diagonal and off-diagonal cells are needed: the operations of these two types of PEs are shown in detail in Fig. 4(b). The vectoring cell and rotation cell will be introduced with the Givens rotation description. There is a slight difference between the off-diagonal cells in the upper-triangle part and those in the lower-triangle part. Fig. 4(b) shows only the off-diagonal cell in the upper-triangle part. Those off-diagonal cells in the lower-triangle part have $y_{in}$ and $c_{in}$ come from the top, while $c_{out}$ leaves from the bottom. Except for this minor difference in the data interface, the operations are the same as the off-diagonal cells in the upper-triangle part. Additionally, in Fig. 4(b) the dotted lines represent the logic control signals transmitted between cells, and the solid lines represent the data transmitted. To initialize the process, each element of the matrices $\mathbf{R}$ and $\mathbf{Q}^H$ (denoted as $r$ and $q$, respectively, in Fig. 4(b)) from QR decomposition are stored in the PE at the corresponding position. For example, $q_{i,i}$ and $r_{i,i}$ are stored in the corresponding diagonal cell $D_{ii}$. The off-diagonal elements $q_{i,j}$ and $r_{i,j}$ are stored in the off-diagonal cell $O_{ij}$. Additionally, the elements of the unimodular matrix $\mathbf{T}$ (denoted as $t$ in Fig. 4(b)) are also stored in the arrays, with $\mathbf{T}$ initially set to the identity matrix.

Fig. 5 shows the overall processes of the full size reduction in the systolic array. In this stage, two major processing modes are defined in each diagonal and off-diagonal cell, the *size reduction mode* and the *data mode* as detailed in Fig. 4(b). In the *size reduction mode*, the objective of each cell is to make condition (9) valid. On the other hand, the cell only performs data propagation in the *data mode*. The cell decides to work in either mode depending on the occurrence of the logic control signal "#". For simplicity, we assume the cells execute all operations in the *data mode* or the *size-reduction mode* in one normalized cycle[2]. At $T = 0$, the external controller sends in the logic control signal "#" to cell $D_{33}$ through cell $D_{44}$. At $T = 1$, cell $D_{33}$ works in the *data mode* due to the control signal "#" and spreads out the "#" logic control signal to the neighboring 3 cells. Meanwhile, $D_{33}$ sends out the data $(r_{3,3}, t_{3,3})^{(*)}$ to cell $O_{34}$. Note that the superscript (*) is a tag bit attached to the data, which indicates that the data are sent out by a diagonal cell. The occurrence of a tag bit (*) will drive the off-diagonal cell to compute $\mu$, and use $\mu$ to update the data stored in that cell. As a result, at $T = 2$, cell $O_{34}$ sends out the newly computed $\mu$ to the two neighboring cells $O_{24}$ and $D_{44}$. At next time instant ($T = 3$), the $\mu$ signal generated by $O_{34}$ meets the data coming from cell $O_{23}$ ($O_{43}$) inside the cell $O_{24}$($D_{44}$), and executes the size reduction update. At the same time instant, data $(r_{2,2}, t_{2,2})^{(*)}$ enter cell $O_{23}$. As cell $O_{34}$ did at $T = 2$, cell $O_{23}$ computes $\mu$, updates $(r_{2,3}, t_{2,3})$, and sends out $\mu$ to the neighboring cells $O_{13}$ and $D_{33}$. The most important fact here is that cell $O_{23}$ also propagates the data $(r_{2,2}, t_{2,2})^{(*)}$ to cell $O_{24}$, and thus starts the column operations between column 2 and column 4 at $T = 4$. Similarly, the column operations between column 1 and column 4 begins at $T = 6$ as $(r_{2,2}, t_{2,2})^{(*)}$ enter cell $O_{14}$. Essentially, full size reduction is a series of column operations between column $j$ and columns $j-1, j-2, \cdots, 1$, for all $2 \leq j \leq m$, and we

[2]The real hardware cycle counts could be multiples of the normalized cycle.

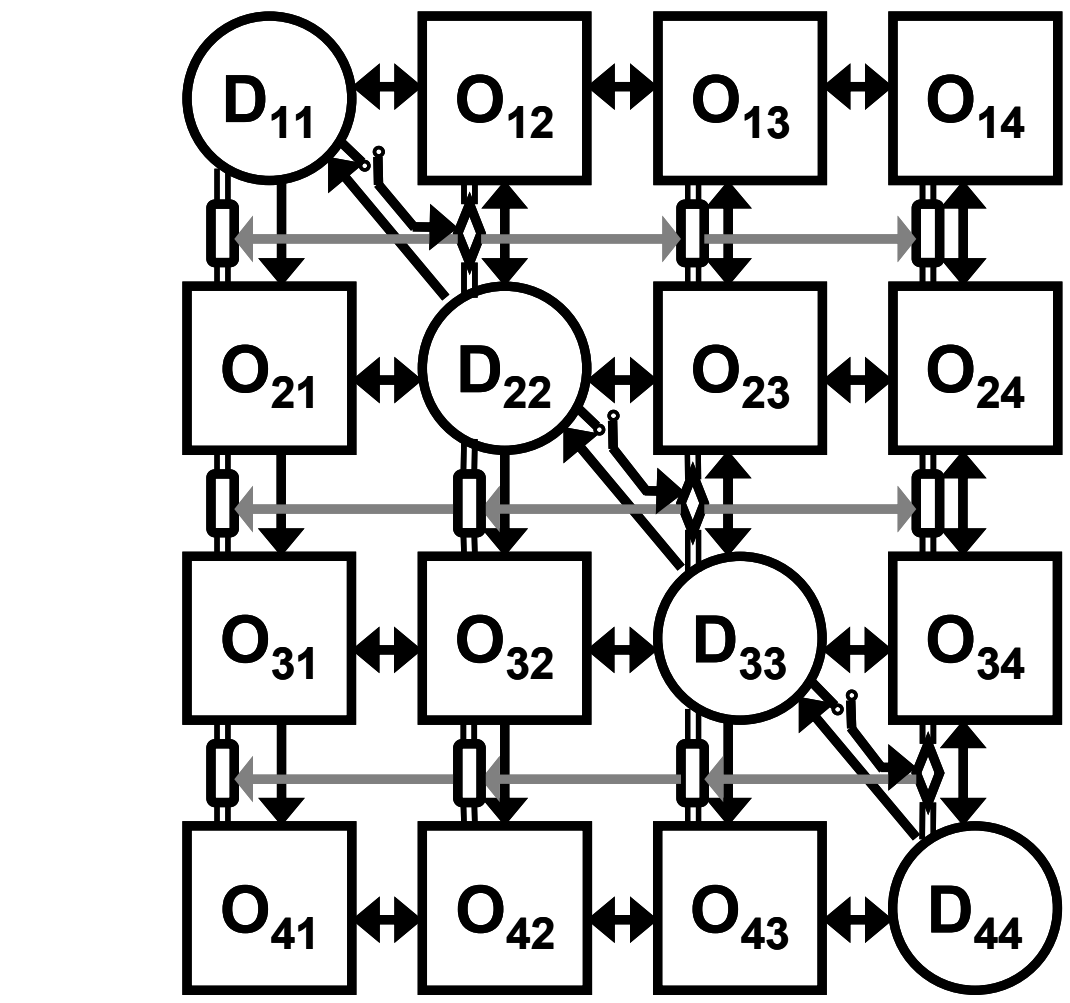

(a)

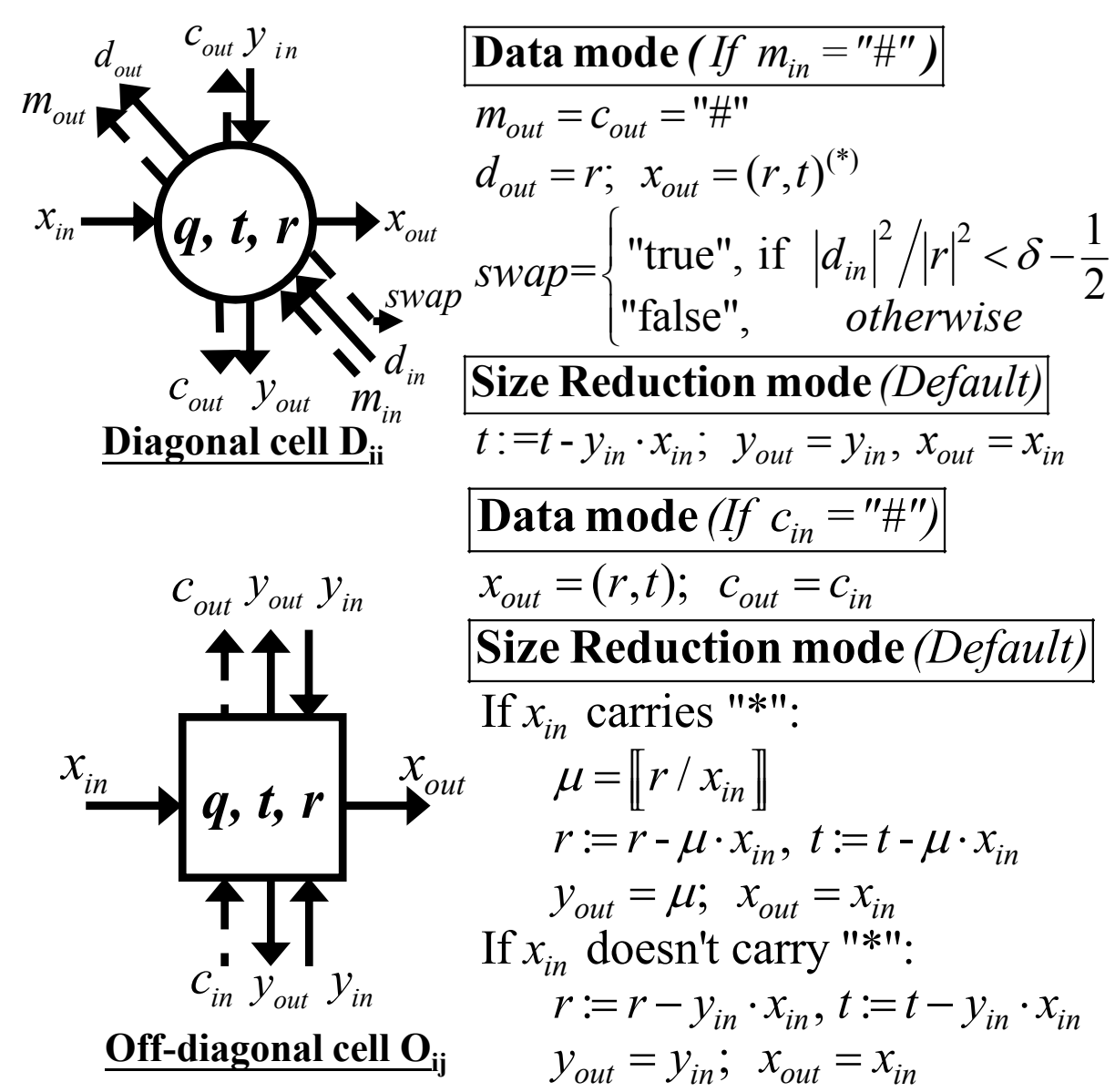

(b)

Fig. 4. (a) The systolic array for the linear LRAD of $4 \times 4$ MIMO system. (b) The operations of diagonal and off-diagonal cells in the systolic array. ("*" is an indicator bit used to control the flow of the algorithm, as explained in Section IV-A)

can conclude the following facts for an $m \times m$ MIMO system:

**[Fact 1]** In this systolic flow, the column operation between column $j$ and column $i$ $(i < j)$ begins at $T = m + j - 2i$ as $(r_{i,i}, t_{i,i})^{(*)}$ enters cell $O_{ij}$.

*Proof:* Data $(r_{i,i}, t_{i,i})^{(*)}$ leaves cell $D_{ii}$ at $T = m - i$, and it takes $j - i$ cycles to have $(r_{i,i}, t_{i,i})^{(*)}$ propagates from cell $D_{ii}$ to cell $O_{ij}$. ■

**[Fact 2]** All column operations on column $j$ end at $T = 2m + j - 3$ in cell $O_{mj}$.

*Proof:* In this systolic flow, the last column operation on column $j$ is always between column $j$ and column 1, which starts at $T = m + j - 2$ in cell $O_{1j}$ according to fact 1. It takes $m - 1$ more cycles to propagate $\mu$ from cell $O_{1j}$ to cell $O_{mj}$ and finish the column operation. ■

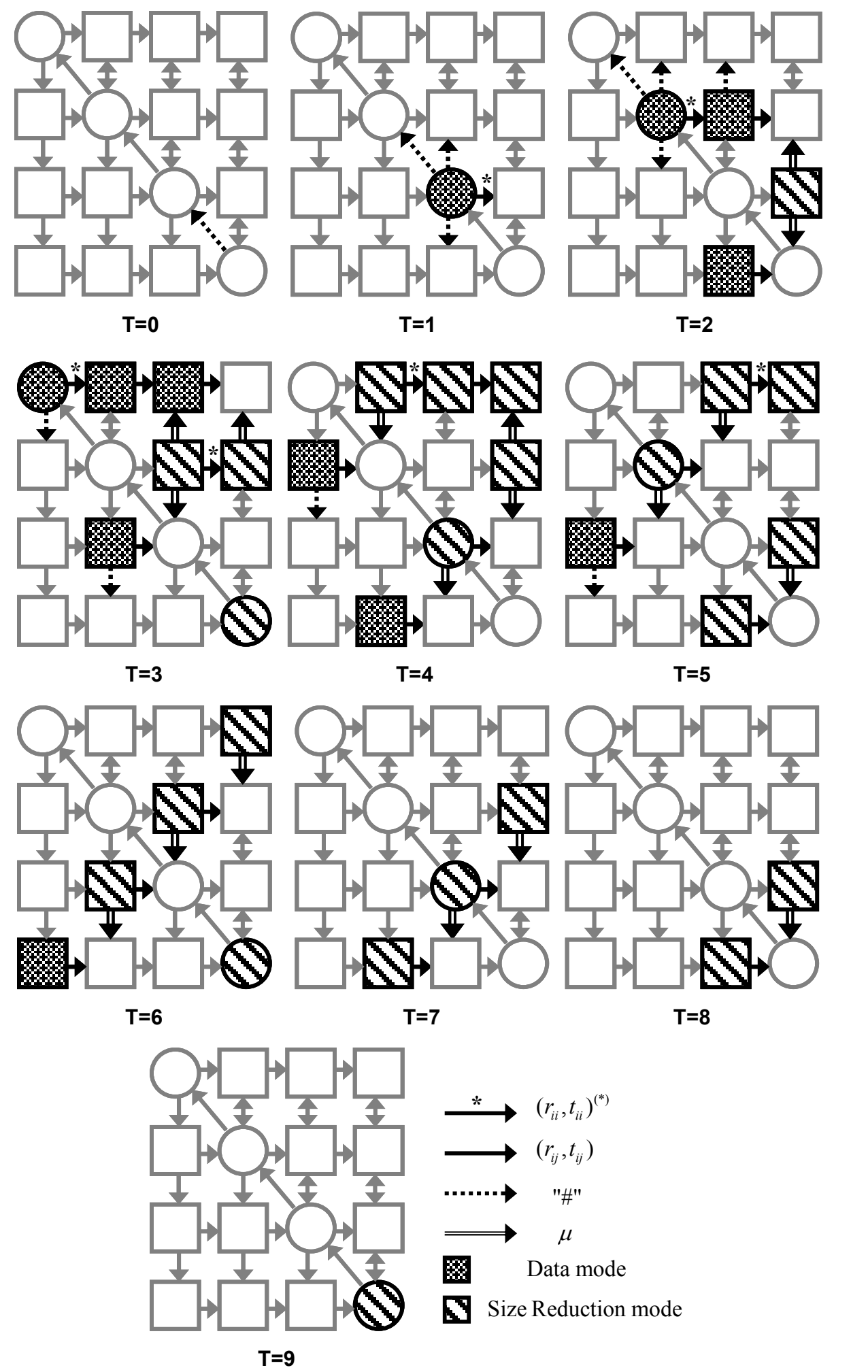

Fig. 5. Flow chart of the full size reduction operations in the systolic array.

**[Fact 3]** The full size reduction ends at $T = 3m - 3$, when all updates on column $m$ are done.

*Proof:* The full size reduction ends when column $m$ finish all the column operations. Therefore, it follows the result in fact 2 that the last step is at $T = 3m - 3$. ■

Referring back to the example mentioned in Section III-A, we can have a more concrete view about the advantage of FSR-LLL over the conventional LLL form when a systolic array is used. If FSR-LLL is applied, the systolic array takes a total of $3m - 3$ cycles to end the all processes. However, for non-systolic LLL, it takes $2m + j - 3$ to process column $j$, and all column operations cannot be done in parallel. So the total time to perform size reduction in non-systolic LLL would be $\sum_{j=3}^{m} (2m + j - 3) = 2.5m^2 - 6.5m + 3$ cycles in that example. In this case, as $m$ increases beyond 3, the advantage of FSR-LLL over the conventional format becomes significant.

*2) Givens Rotation:* As mentioned in Section III-C, we use Siegel condition in the lattice reduction algorithm, which only relates two $r$ elements in the neighboring diagonal cells. Hence, this condition can be checked during a full size reduction step. For example, in Fig. 5 at $T = 1$, cell $D_{33}$ sends data $r_{3,3}$ to cell $D_{22}$ along with the "#" signal. At the next time instant, cell $D_{22}$ will check this condition based on $|r_{3,3}|^2 \big/ |r_{2,2}|^2$, and also generate the logic control signal "*swap*" (see Fig. 4(b)). If $\delta - 1/2$ is greater than $|r_{i,i}|^2 \big/ |r_{i-1,i-1}|^2$ then "*swap*" is "*true*", and drives the vectoring cell to work. The operations of vectoring and rotation cells are shown in Fig. 6. The vectoring cell zeros out the input data $\beta$ by the Givens rotation matrix $\mathbf{G}$, which is calculated based on Table I lines 13 to 15. The rotation cell simply rotates the input data with the angle $\Theta$ given by the neighboring vectoring cell. Hence, the vectoring and rotation cells also work in a systolic way, with the rotation angle $\Theta$ propagating between cells. As shown in Fig. 4(a), there are 3 rotation cells and 1 vectoring cell between every two consecutive rows of the systolic array. These cells perform the Givens rotation to the $\mathbf{R}$ and $\mathbf{Q}^H$ data in those two rows. The vectoring cell is located between cells $D_{ii}$ and $O_{i-1,i}$ because the Givens rotation step is executed prior to the column-swap step in FSR-LLL, and data $r_{i,i}$ need be zeroed so that the matrix $\mathbf{R}$ is still upper triangular after column swap.

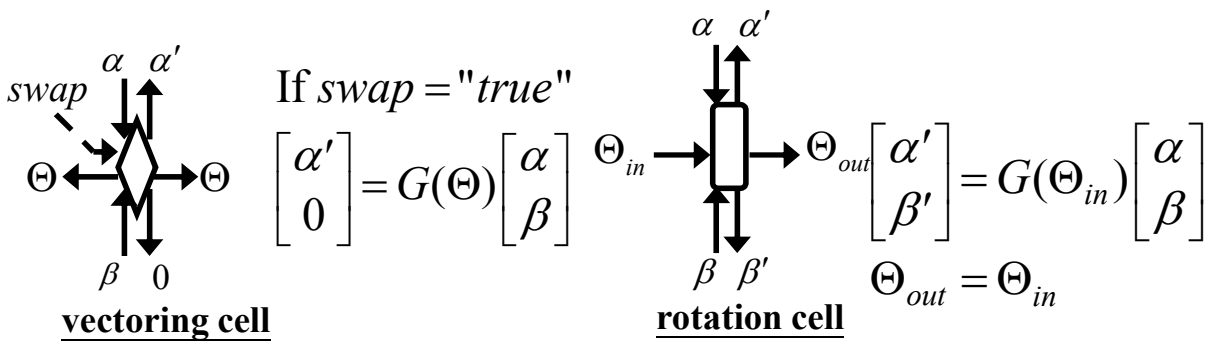

Fig. 6. The operations of vectoring cells and rotation cells in the systolic array.

Note that Givens rotation only applies to rows $k'$ and $k'-1$ during one iteration of FSR-LLL if $k'$ exists (lines 13~16 in Table I). However, every $D_{ii}$ ($i = 1, \cdots, m-1$) could generate the "*swap*" signal during the full size reduction step. Therefore, we need a direct access from the external controller to each diagonal cell in order to control the data path between the diagonal cell and the vectoring cell. Namely, only cell $D_{k'k'}$ can pass the signal "*swap*" to the vectoring cell and perform the Givens rotation to rows $k'$ and $k'-1$. In Fig. 4(a), we use a "*switch*" symbol between each pair of a diagonal cell and a vectoring cell to represent the control by the external controller. Only one switch is turned on during one iteration.

Additionally, a Givens rotation on rows $k'$ and $k'-1$ can begin right after $r_{k'-1,k'}$ is updated during the full size reduction step. For example, $r_{3,4}$ is updated at $T=2$ as shown in Fig. 5, and Givens rotation on rows 3 and 4 could start as early as $T = 3$ without any interference to the remaining operations of full size reduction. This way, the time necessary to perform Givens rotations can be partially hidden by the full size reduction and this is the reason why we want the Givens rotation to occur prior to column swap in our design. For hardware implementation, one could consider using only one rotation cell between every two neighboring rows or the systolic array to reduce the hardware complexity. This will not lead to significant increase in time if we consider performing Givens rotation and full size reduction in parallel.

*3) Column swap:* The columns $k'$ and $k'-1$ of $\mathbf{R}$ (and $\mathbf{T}$) should be swapped, after the Givens rotation is done. However, it is possible that the column swap be partially overlapped in time with size reduction and Givens rotation. For example, the column swap could begin after $\mathbf{R}$ being rotated but prior to $\mathbf{Q}^H$ being updated since there is no need to swap columns of $\mathbf{Q}^H$.

The FSR-LLL stops when there is no possible column swap, i.e., a $k'$ in Table I, line 11, does not exist. The system flow (lines 3, 18 and 20 in Table I) is controlled by the external processor. The lattice reduced matrices $\tilde{\mathbf{R}}$ and $\tilde{\mathbf{Q}}^H$ and the unimodular matrix $\mathbf{T}$ stay in the PEs. The systolic array along with these matrices will be used for linear detection, as described in Section V below.

### B. All-Swap Lattice Reduction (ASLR) Algorithm

The ASLR algorithm can also be performed by the systolic array shown in Fig. 4(a). The process of full size reduction is the same as in Fig. 5. During full size reduction, the Siegel condition is also checked in each diagonal cell $D_{11}$~$D_{m-1,m-1}$. If the current value of "*order*" is even (odd), then the "*switch*" between each cell $D_{k-1,k-1}$ with even (odd) index $k$ and the vectoring cell is turned on by the external controller. Consequently, for every even (odd) index $k$, Givens rotation between rows $k-1$ and $k$ could be executed if needed. As for the column swap step, more than one pair of columns could be swapped during one iteration, but all these pairs are swapped in parallel. Hence, the time spent on columns swap is the same as on swapping a single pair of columns. Based on this observation, we can expect the systolic ASLR to work more efficient than the systolic FSR-LLL. Comparisons between these two algorithms in terms of bit-error-rate performance and of efficiency in execution time are deferred to the next subsection.

Note that in our description we limit the applications of this systolic array only to an $m \times m$ MIMO system. For $m \times m$ MMSE-LRAD, although the matrix $\mathbf{Q}^H$ is $m \times 2m$ (the extended channel model in (5)), we can treat the submatrix $\mathbf{Q}^H_{1:m,(m+1):2m}$ as another square matrix, and store each element of $\mathbf{Q}^H_{1:m,(m+1):2m}$ in the PE at the corresponding position. Namely, $q_{i,j}$ and $q_{i,j+m}$ should be stored in the same PE, which still keeps the systolic array square.

### C. Comparison between FSR-LLL and ASLR algorithm

First, we compare the two algorithms in term of bit-error-rate (BER) performance, and also compare them with the conventional LLL algorithm. In our simulation, 4-QAM is assumed for the transmitted symbols. The constant $\delta$ is set to 0.99 in all algorithms for fair comparison. Let $E_b$ be defined as the equivalent energy per bit at the receiver, and thus $E_b/N_0$ is $m/(\sigma^2 \log_2 M)$. The Fig. 7(a) shows the BER results of minimum mean-square-error LRAD (in $4 \times 4$ and $8 \times 8$ MIMO systems) based on FSR-LLL (denoted as MMSE-FSR), ASLR algorithm (denoted as MMSE-ASLR) and the LLL algorithm (denoted as MMSE-LLL). The BER results for ML detection and MMSE without lattice reduction are also shown for comparison. As $\delta = 0.99$, the FSR-LLL and ASLR work as well as LLL algorithm, and even slightly better in the case of $m = 8$. It clearly shows that using the insignificantly weaker Siegel condition does not deteriorate the BER performance of linear detections in an MIMO system as compared to the conventional LLL. In Fig. 7(b), the BER performance of an $4 \times 4$ MIMO system using LR-aided MMSE SIC based on different lattice reduction algorithms are shown. Unlike the linear detection case, the LLL-aided SIC works better than the other two algorithms. Since the detection of the first layer in SIC dominates the overall performance, it implies that due to Siegel condition the FSR-LLL-reduced or the ASLR-reduced channel provides lower SNR for the first layer in SIC than the one given by the conventional LLL. Additionally, FSR-LLL and ASLR lead to almost the same results in all three MIMO systems, which is consistent with the results in Fig. 3. Hence, we can conclude that although FSR-LLL and ASLR give different lattice reduced matrices, the LRAD based on these two algorithms have very similar BER performance.

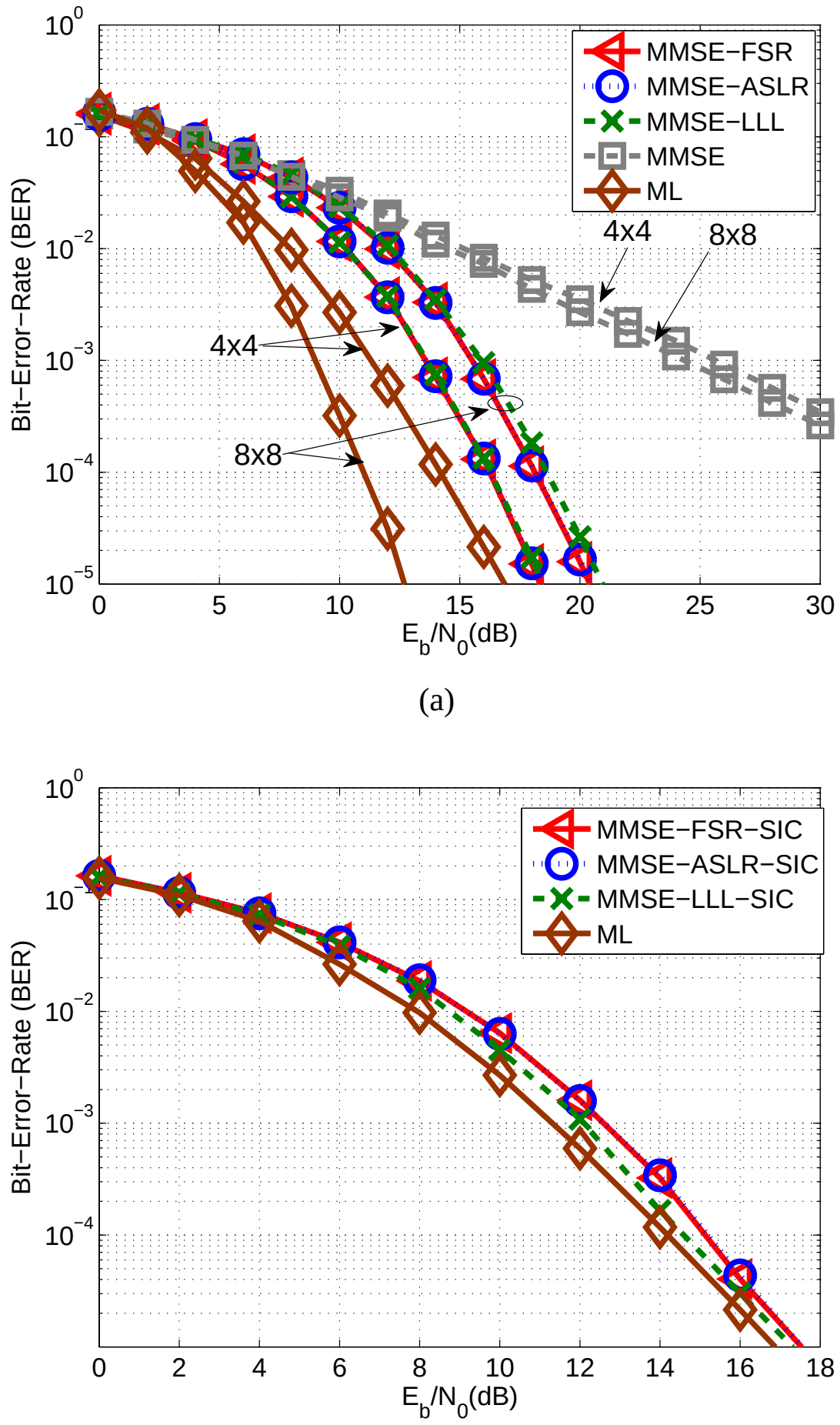

Fig. 7. BER performance of FSR-LLL and ASLR- based MMSE LRAD. (a)Linear detection ($4\times4$ and $8\times8$ MIMO systems) (b)SIC (an $4\times4$ MIMO systems)

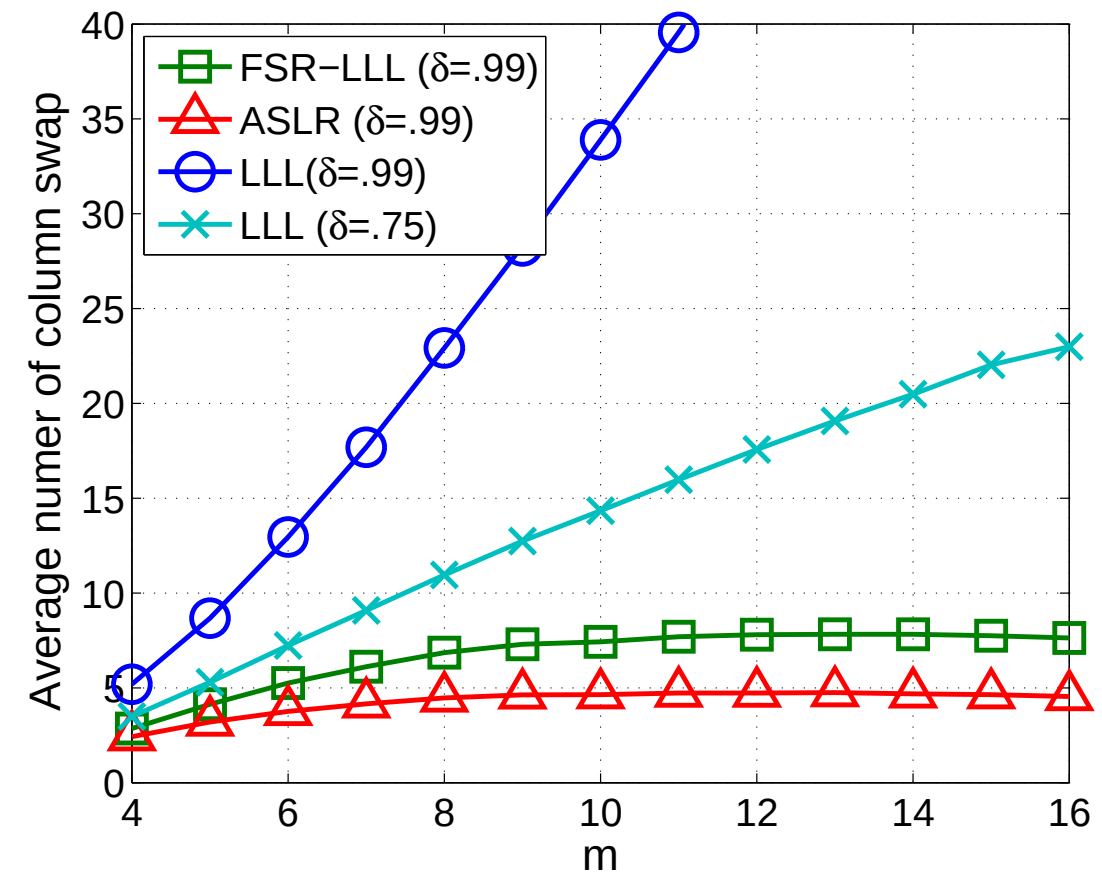

Fig. 8. The average number of column swaps in FSR-LLL, ASLR and LLL-aided MMSE detection in $m \times m$ MIMO system with $E_b/N_0$ fixed at 20 dB.

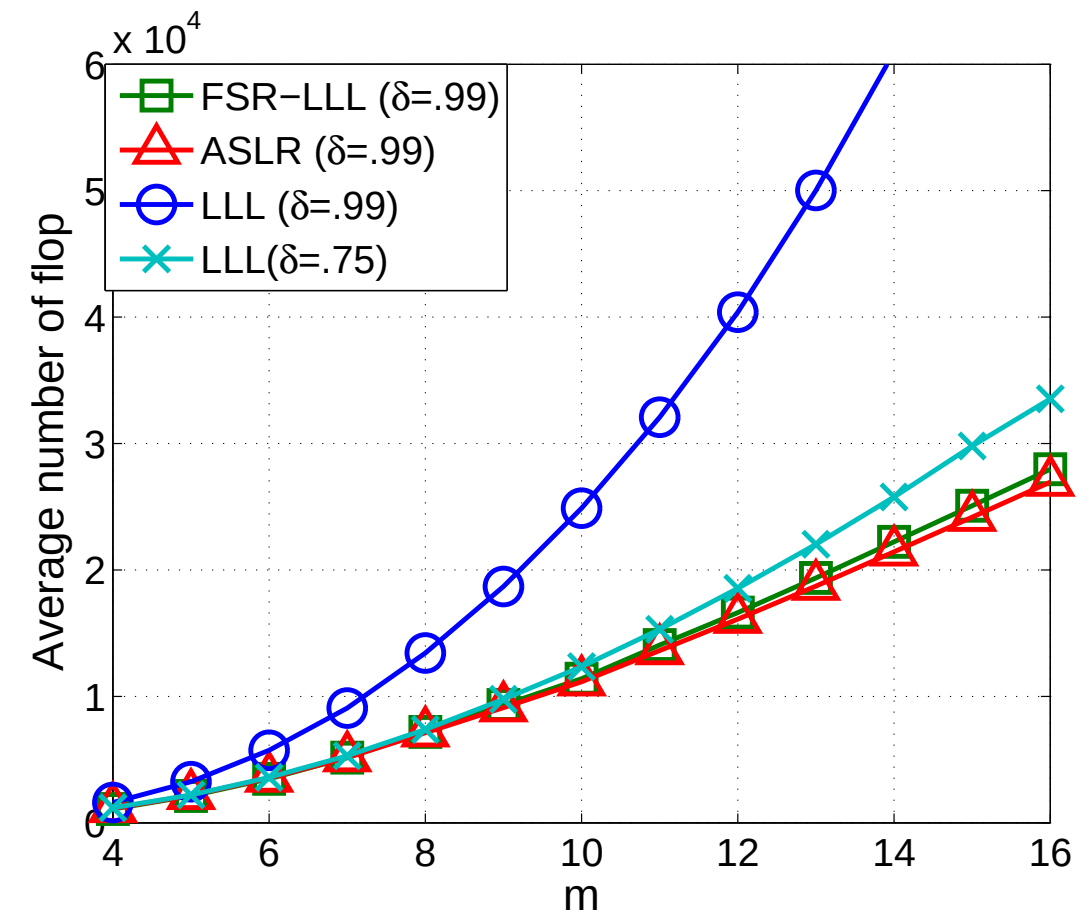

Fig. 9. The average number of floating point operations in FSR-LLL, ASLR and LLL-aided MMSE detection in $m\times m$ MIMO system with $E_b/N_0$ fixed at 20 dB.

Next, we compare the efficiency of the systolic array for both algorithms. It is known that the number of iterations of FSR-LLL and ASLR depends on the condition number of the channel matrix. If $\mathbf{H}$ is well-conditioned, lattice reduction takes less iterations, and thus less cycles in the systolic array. Since both algorithms begin with full size reduction, the total execution time is fully determined by the number of column swaps in the overall process. Less column swapping implies less iterations. Fig. 8 shows the average number of column swaps in FSR-LLL and ASLR-aided MMSE detection (with $E_b/N_0$ fixed at 20dB) in $m\times m$ MIMO systems ($m = 4$~16). Note that for ASLR we count all the even or odd columns swaps during one iteration as only one swap since they are executed in parallel. In an $4\times4$ MIMO, the difference between the two algorithms is almost negligible. However, as the number of antennas grows, the advantage of ASLR becomes significant. For $m \geq 8$, ASLR has less than 65% the column swaps comparing to FSR-LLL. Based on BER performance and time-efficiency comparisons, ASLR should be a better algorithm to be applied on our systolic array, especially with a large number of antennas.

For comparison, the results of the conventional LLL with $\delta = 0.99$ and 0.75 are also shown in Fig. 8. As expected,

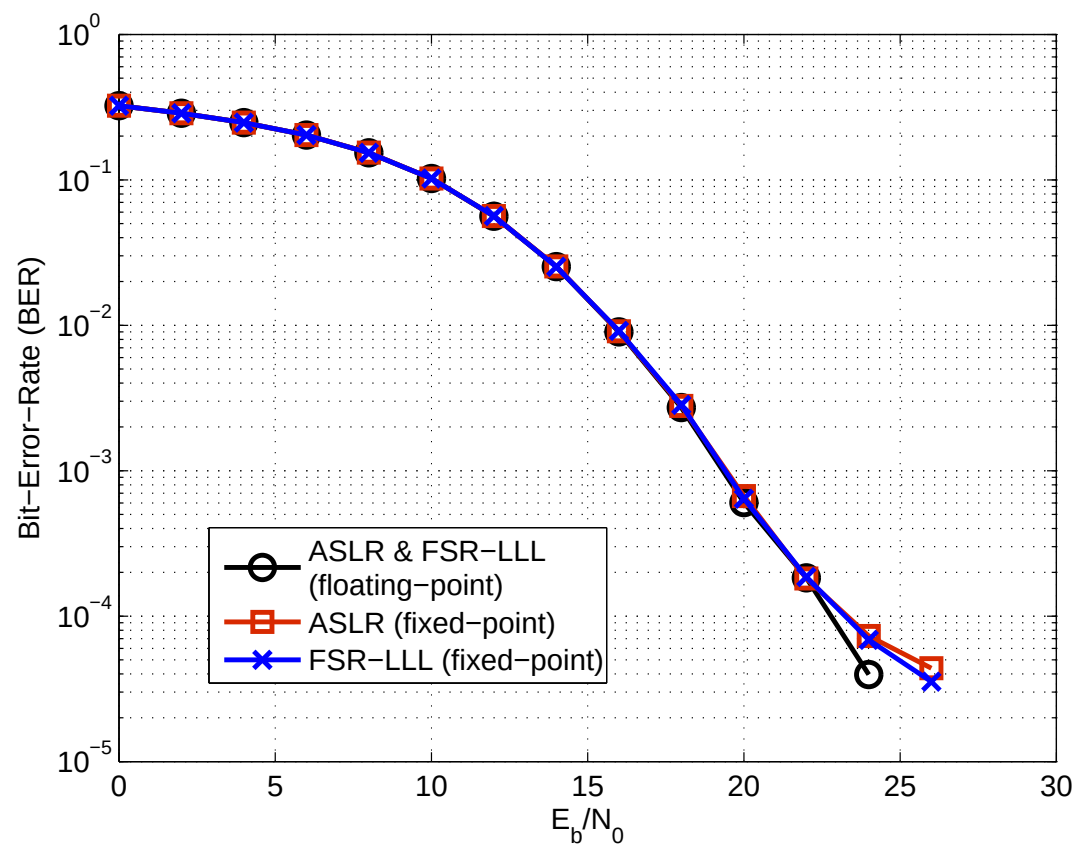

Fig. 10. Comparison between the fixed-point and floating-point lattice reduction algorithms using ZF-SIC in an $4 \times 4$ MIMO system

TABLE III
FPGA IMPLEMENTATION RESULTS

<table>
<tr><td>Target Algorithm</td><td colspan="2">ASLR</td><td colspan="2">FSR-LLL</td><td colspan="2">CLLL [14]</td></tr>
<tr><td>Device</td><td>Virtex 5[1]</td><td>Virtex 6[2]</td><td>Virtex 5[1]</td><td>Virtex 6[2]</td><td>Virtex 4</td><td>Virtex 5</td></tr>
<tr><td>Slices</td><td>2322 /20480</td><td>1812 /20000</td><td>2335 /20480</td><td>1798 /20000</td><td>3617 /67584</td><td>1712 /17280</td></tr>
<tr><td>Clock Frequency</td><td>160MHz</td><td>249MHz</td><td>155MHz</td><td>247MHz</td><td>140 MHz</td><td>163 MHz</td></tr>
<tr><td rowspan="4">Avg. cycles(time) per channel matrix</td><td colspan="2">80 (SQRD)</td><td colspan="2">84 (SQRD)</td><td colspan="2" rowspan="2">130 (SQRD)</td></tr>
<tr><td>500.0ns</td><td>321.3ns</td><td>541.9ns</td><td>340.1ns</td></tr>
<tr><td colspan="2">146 (QRD)</td><td colspan="2">164 (QRD)</td><td rowspan="2">928.6ns</td><td rowspan="2">797.5ns</td></tr>
<tr><td>912.5ns</td><td>586.3ns</td><td>1058.1ns</td><td>664.0ns</td></tr>
</table>

[1]part number: XC5VFX130T [2]part number: XC6VLX130T

LLL with $\delta = 0.99$ has a higher complexity than LLL with $\delta = 0.75$. Furthermore, the conventional LLL has a much higher average number of column swaps than FSR-LLL and ASLR have in the higher-dimensional MIMO system ($m \geq 8$). However, it is not fair to conclude that the complexities of FSR-LLL and ASLR are much lower than the conventional LLL; in fact, full size reductions are performed in the former two algorithms, and full size reduction needs more computation efforts than the conventional size reduction in LLL. In Fig. 9, we compare the number of floating point operations (flop) in LLL, FSR-LLL, and ASLR using the same settings as in Fig. 8. The flops are counted in terms of number of real additions and real multiplications. One complex addition is counted as two flops (two real additions) and one complex multiplication is counted as six flops (four real multiplications and two real additions). The complexity of QR decomposition is neglected, since this is done only once at the beginning of the three algorithms. It is shown that LLL with $\delta = 0.99$ has the highest complexity among the three. Under the same $\delta$ $(= 0.99)$ setting, FSR-LLL and ASLR have a much lower computational complexity than LLL. On the other hand, the complexity of LLL with $\delta = 0.75$ is just slightly higher than FSR-LLL and ASLR, even though the average number of column swaps of LLL with $\delta = 0.75$ is more than two times larger than the one of ASLR for $m \geq 10$. This implies that the process of full size reduction introduces some additional complexity. However, thanks to the (insignificantly) weaker Siegel condition, the complexities of ASLR and FSR-LLL for $m \geq 10$ are less than 50% of the complexity of LLL with the same $\delta$ setting.

To further explore the advantage of using systolic array, we implement our proposed architecture for an $4 \times 4$ MIMO system onto FPGA. We performed our design using Xilinx System Generator 11.5 (XSG) block-set in the Simulink design environment. A Verilog Hardware Description Language (HDL) code is then generated automatically by XSG and is synthesized by Xilinx XST. The place and route is done by Xilinx ISE 11.5. The word-length of $\mathbf{R}$, $\mathbf{Q}^H$, $\mathbf{T}$ and $\mu$ are set to (18,13), (14,13), (8,0) and (3,0), respectively. As mentioned in Section III-C, the division in Siegel condition check can be avoided by using a comparator. The divisions in the Givens rotation are implemented by the Newton-Raphson iterative algorithm [37]. As for $\mu$, it can be easily shown by simulation that $|\mu|$ is either 0, 1, or 2 over 99.7% of the time. Hence, we can simply use a set of comparators to determine the value of $\mu$ instead of using a division. For those $|\mu|$ greater than 2 are saturated to 2, which rarely happened. The BER performance of the fixed-point systolic implementation for an $4 \times 4$ MIMO system is shown in Fig. 10, where 16-QAM modulation and ZF-SIC detection are applied. The implementation results are shown in Table III. We consider both QRD and SQRD as the pre-processes of the lattice reduction algorithms. From the results, ASLR is superior to FSR-LLL in terms of the average processing time, and this advantage is significant when QRD is applied. The hardware complexity for ASLR and FSR-LLL are about the same, since they only differ from each other in the external controllers. It is also clear that SQRD reduces the average processing time by over 45% comparing to using the normal QRD, at the cost of higher computation efforts on SQRD.

In Table III, the FPGA implementation result for the conventional complex LLL (CLLL) [14] is also listed for comparison. Under Virtex 5 and with SQRD, systolic ASLR operates at a slightly lower speed than the one of CLLL; however our designs require only 61.5% average clock cycles of theirs. As a result, ASLR is on average faster than CLLL by a factor of 1.6. This verifies the high-throughput advantage of the systolic arrays. On the other hand, systolic arrays implementation may have higher hardware complexity since it requires several processing elements to work in parallel. The results in Table III shows that our designs occupied 36~38% more FPGA slices than the one in CLLL. However, as the fast the advance of FPGA technology and the semiconductor processing, one may consider to trade some areas for a faster processing speed. As shown in Table III, when using the latest Xilinx Virtex 6 FPGA device, our systolic designs could run up to 249MHz and it only requires less than 10% of the total FPGA slices.

## V. SYSTOLIC ARRAY FOR DETECTION METHODS

### A. Linear Detection in Systolic Array

After lattice reduction, the matrices $\tilde{\mathbf{Q}}^H$ and $\tilde{\mathbf{R}}$, along with the unimodular matrix $\mathbf{T}$, are stored in the systolic array. As shown in Fig. 2, the first step of a linear detection consists of premultiplying the received signal vector $\mathbf{y}$ by $\tilde{\mathbf{H}}^\dagger$, which yields $\hat{\mathbf{x}} = \tilde{\mathbf{H}}^\dagger \mathbf{y} = \tilde{\mathbf{R}}^{-1}\tilde{\mathbf{Q}}^H\mathbf{y}$. Second, the result of a matrix–vector multiplication needs to be rounded element-wise. The final step is to multiply the rounded results by the unimodular matrix $\mathbf{T}$ and constrain all results within the constellation boundary. If $\hat{\mathbf{x}}_q$ denotes the element-wise-rounded $\hat{\mathbf{x}}$, the final decision of the LRAD is $\hat{\mathbf{x}}_{LR} = \mathcal{Q}(\mathbf{T}\cdot\hat{\mathbf{x}}_q)$, as described in Section II-C.

In the following discussion, we assume an $4\times 4$ MIMO system, and consider the zero-forcing detection first. The first and last steps of a linear detection can be implemented by the same systolic array of Fig. 4 without using extra cells. As for the rounding and the final constellation boundary check, they should be done outside the systolic array (they are not shown in Fig. 11). To execute $\hat{\mathbf{x}} = \tilde{\mathbf{R}}^{-1}\tilde{\mathbf{Q}}^H\mathbf{y}$ in the systolic array, we separate it into two matrix–vector multiplications $\mathbf{v} = \tilde{\mathbf{Q}}^H\mathbf{y}$ and then $\hat{\mathbf{x}} = \tilde{\mathbf{R}}^{-1}\mathbf{v}$. Since $\tilde{\mathbf{Q}}^H$ stays in the systolic arrays after the lattice reduction ends, the received signal vector $\mathbf{y}$ can be fed to the systolic arrays from the top in a skewed manner as shown in Fig. 11(a). The vector $\tilde{\mathbf{Q}}^H\mathbf{y}$ is pumped out from the rightmost column of the array. Diagonal and off-diagonal cells are needed at this stage, and the operations of the cells are shown in Fig. 12(a). Every cell performs the multiply-and-add operation. If MMSE is chosen, the input vector should be changed to an $2m\times 1$ vector $\underline{\mathbf{y}}$ according to the extended model (5). Let $\underline{\mathbf{y}} = \left[\mathbf{y_1^T}\ \mathbf{y_2^T}\right]^T$ and $\tilde{\mathbf{Q}}^H = [\mathbf{Q_1}\ \mathbf{Q_2}]$ , where $\mathbf{y}_1$, $\mathbf{y}_2$ are $m\times 1$ vectors and $\mathbf{Q}_1$, $\mathbf{Q}_2$ are $m\times m$ matrices. As mentioned in Section IV-B, the elements of $\mathbf{Q}_1$ and $\mathbf{Q}_2$ are stored in the same PEs. To compute $\mathbf{v} = \tilde{\mathbf{Q}}^H\underline{\mathbf{y}}$ using the systolic array, first we let $\mathbf{y}_1$ enter the array from the top and multiply it by $\mathbf{Q}_1$, which is the same as shown in Fig. 11(a). Then $\mathbf{y}_2$ enters the array right after $\mathbf{y}_1$, also in a skewed manner, and is multiplied by $\mathbf{Q}_2$. Hence, for MMSE we need an extra operation at the output of the array, which is $\mathbf{v} = \mathbf{Q}_1\mathbf{y}_1 + \mathbf{Q}_2\mathbf{y}_2$. For the remaining operations in the systolic array, there is no difference between ZF and MMSE detections.

The second stage consists of computing $\hat{\mathbf{x}} = \tilde{\mathbf{R}}^{-1}\mathbf{v}$. Instead of computing $\tilde{\mathbf{R}}^{-1}$ directly, the following recursive equation [38] is considered for the systolic design

$$\hat{x}_j = \frac{1}{\tilde{r}_{j,j}}\left(v_j - \sum_{i=j+1}^{m}\tilde{r}_{j,i}\hat{x}_i\right),\ j \text{ starts from } m \text{ to } 1. \quad (14)$$

According to (14), it is clear that $\tilde{\mathbf{R}}^{-1}\mathbf{v}$ can be computed directly from the components of $\tilde{\mathbf{R}}$ without computing $\tilde{\mathbf{R}}^{-1}$. Additionally, it can be implemented by the upper triangle part of the systolic array, where matrix $\tilde{\mathbf{R}}$ has already been stored. As shown in Fig. 11(b), the vector $\mathbf{v} = \tilde{\mathbf{Q}}^H\mathbf{y}$ enters the array from the right, and $\hat{\mathbf{x}} = \tilde{\mathbf{R}}^{-1}\mathbf{v}$ is computed by the triangular array with cell operations shown in Fig. 12(b). The output vector $\hat{\mathbf{x}}$ is then rounded element-wise outside the systolic array. The final step consists of multiplying the quantized vector $\hat{\mathbf{x}}_q$ by the unimodular matrix $\mathbf{T}$, which is also stored in the array. Similar to the first step of a linear detection, it is a matrix–vector multiplication between $\mathbf{T}$ and $\hat{\mathbf{x}}_q$. Hence, the data flow in Fig. 11(c) is the same as Fig. 11(a). The cell

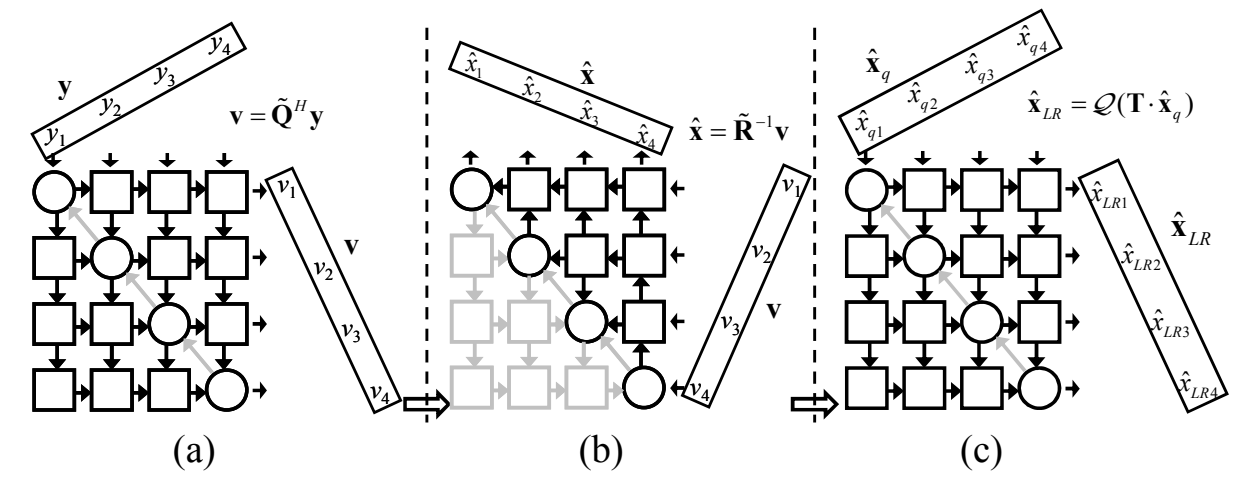

Fig. 11. The linear detection operations in the systolic array. (a)$\mathbf{v} = \tilde{\mathbf{Q}}^H\mathbf{y}$ (b)$\hat{\mathbf{x}} = \tilde{\mathbf{R}}^{-1}\mathbf{v}$ (c)$\hat{\mathbf{x}}_{LR} = \mathcal{Q}(\mathbf{T}\cdot\hat{\mathbf{x}}_q)$.

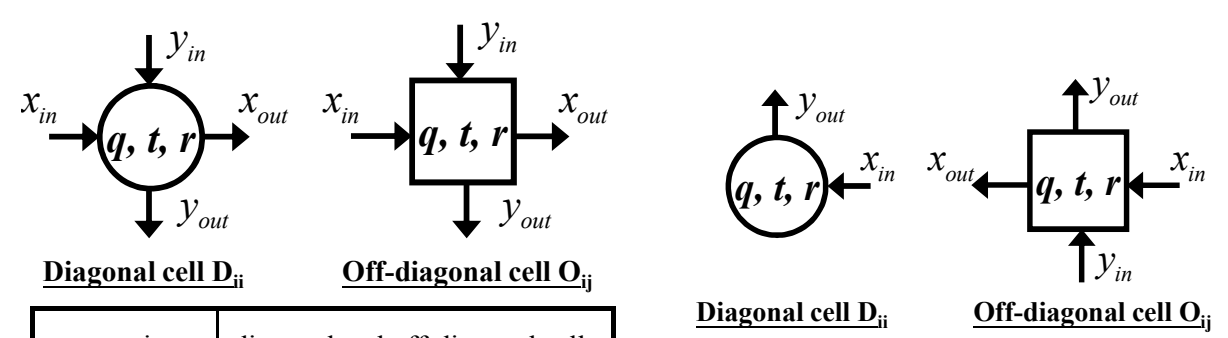

| operation | diagonal and off-diagonal cells |
|---|---|
| $\tilde{\mathbf{Q}}^H\mathbf{y}$ | $x_{out} = x_{in} + q\cdot y_{in};\ y_{out} = y_{in}$ |
| $\mathbf{T}\cdot\hat{\mathbf{x}}_q$ | $x_{out} = x_{in} + t\cdot y_{in};\ y_{out} = y_{in}$ |

(a)

| operation | diagonal cell | off-diagonal cell |
|---|---|---|
| $\hat{\mathbf{x}} = \tilde{\mathbf{R}}^{-1}\mathbf{v}$ | $y_{out} = x_{in}/r$ | $x_{out} = x_{in} - r\cdot y_{in}$; $y_{out} = y_{in}$ |

(b)

Fig. 12. The detailed operations of the diagonal cells and off-diagonal cells in the systolic array at different stage. (a)$\tilde{\mathbf{Q}}^H\mathbf{y}$ and $\mathbf{T}\cdot\hat{\mathbf{x}}_q$ (b)$\tilde{\mathbf{R}}^{-1}\mathbf{v}$.

operations for $\mathbf{T}\cdot\hat{\mathbf{x}}_q$ are shown in Fig. 12(a), and the array output being quantized to the closest constellation point is the final result $\hat{\mathbf{x}}_{LR}$ of the linear LRAD.

### *B. Spatial-Interference Cancellation in Systolic Array*

The successive spatial-interference cancellation (SIC) can also be performed on this systolic array with some modifications to the PEs. Observing the first step of LR-aided SIC showing in (7), it should be apparent that $\tilde{\mathbf{Q}}^H\mathbf{y}$ can be performed in the systolic array in the same fashion as in Fig. 11(a) and Fig. 12(a). The second step (8) of LR-aided SIC can be done in the systolic array as shown in Fig 13. It is almost the same operations as the one Fig. 12(b), except that we have to do a rounding in the off-diagonal cells $O_{ij}$ at the super-diagonal position ($j = i + 1$). The rounding operations are for the decision of each $\hat{z}_i$. Similar to the linear LRAD, the final step of LR-aided SIC is to multiply $\mathbf{z}$ by the unimodular matrix $\mathbf{T}$ and bound all the output within the QAM constellation. It can be done in the same way as in Fig 11(c) and Fig. 12(a), with $\hat{\mathbf{x}}_q$ being replaced by $\hat{\mathbf{z}}$.

Notice that lattice reduction and linear detection (or SIC) are performed in the same systolic array, and it can be hardware-efficient to share the adder/multiplier/divider designed for lattice reduction processing. For instance, there is one addition, one multiplication, and one division in each diagonal cell, and one addition and one multiplication in each off-diagonal cell for linear detection or SIC, be it ZF or MMSE. These operations are also contained in each cell at the LLL lattice reduction stage. For SIC, it seems that we need extra rounding operations in those off-diagonal cells at the superdiagonal position. Now, we need those rounding operations in the off-diagonal cells during the full size reduction processing as well. Hence, there need be no extra hardware cost (adders or multipliers) in each cell for linear detection. Only extra control logic to the array is needed in order to have each PE work correctly in different modes.

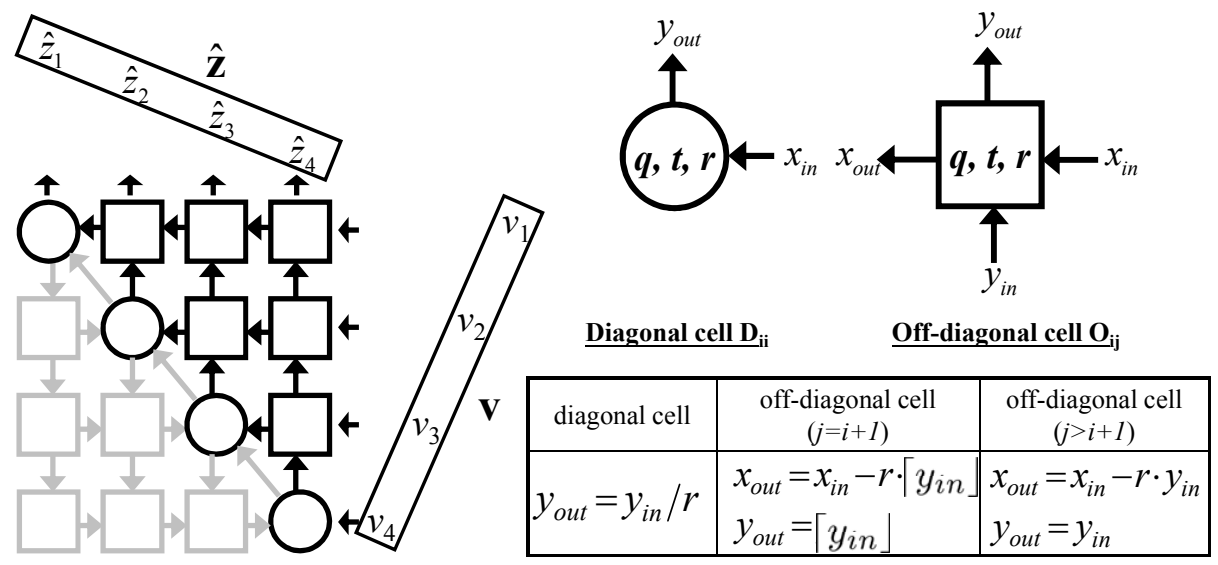

Fig. 13. The data flow and the detailed operations of the cells in the systolic array for the interference-cancellation step of LR-aided SIC.

## VI. Conclusion

In this paper, we have described a systolic array performing LLL-based lattice-reduction-aided detection for MIMO receivers. Lattice reduction and the ensuing linear detection or successive spatial-interference cancellation can be executed by the same array, with minimum global access to each processing element. The proposed systolic array with external logic controller can work with two different lattice-reduction algorithms. One is LLL algorithm with full size reduction, which is a different form of the conventional LLL algorithm and more suitable for parallel processing. The second one is an all-swap complex lattice-reduction algorithm, which generalizes the one originally proposed in [30] for real lattices. Compared to FSR-LLL, ASLR operates on a whole matrix, rather than on its single columns, during the column-swap and Givens-rotation steps. To reduce the complexity of data communications between processing elements in the systolic array, we replace Lovász condition in the LLL algorithm by Siegel condition. Even though Siegel condition is weaker than Lovász condition, the BER performance of LR-aided linear detections based on our two algorithm versions appears to be as good as using the conventional LLL, and the computational complexity is reduced by the relaxation as well. Based on BER performance and time-efficiency comparisons, ASLR should be preferred to FSR-LLL, especially for an MIMO system with a large number of antennas. The FPGA implementation results also show that our proposed systolic architecture for lattice reduction algorithms run about $1.6\times$ faster than the conventional LLL, at the cost of moderate increases of hardware complexity. Additionally, due to the high- throughput property of systolic arrays, our design appears very promising for high-data-rate systems, such as in a MIMO-OFDM system.